# Influence Mechanism Of Environmental Stimulus And Consumer Ethnocentrism On Purchasing Wuliangye: Applications Of Extended Theory Of Planned Behavior (ETPB) And Stimulus-Organism-Response (SOR) Theory

**Ruofeng Rao[1]*, Sarana Photchanachan[2,3]**

[1]*PhD Candidate, Faculty of Management, Shinawatra University, Thailand*
**Corresponding author's email: ruofengrao@163.com*
[2]*Dean, Faculty of Management, Shinawatra University, Thailand,*
*Email: sarana.p@siu.ac.th, https://orcid.org/0000-0001-5916-0250*
[3]*Faculty of Business and Communication, InTi International University, Malaysia*

Environmental stimuli play a pivotal role in triggering impulsive purchases among consumers, while consumers from Sichuan Province, China, exhibit strong ethnocentric tendencies, impacting their decision-making process, particularly regarding Wuliangye liquor, a local product. Through an online survey of 453 Wuliangye consumers from Sichuan, an analysis was conducted using structural equation modeling rooted in the ETPB and SOR theory. This analysis revealed the favorable impact of environmental stimuli and consumer ethnocentrism on purchasing behavior. This influence was found to be partially mediated through perceived value, attitudes, and purchase intention, forming a chain-mediated effect. Notably, purchase intention doesn't always translate to actual buying behavior, with environmental stimuli, consumer ethnocentrism, perceived behavioral control and purchase intention all being robust predictors of purchase behavior. Finally, several management strategies were proposed, aimed at bolstering Wuliangye sales, with a focus on platform development, mid-to-low range product creation, and appealing to Generation Z consumers.

## 1. Introduction

Wuliangye Company Limited, stock code 000858, based in Yibin City, Sichuan Province, China, was established in April 1998. It is one of the largest state-controlled listed companies. The study of Wuliangye's sales is economically significant for China. This paper focuses on Wuliangye consumers in Sichuan Province, especially since the company's main customer base shifted from primarily official groups to individual private consumers following the implementation of the Chinese government's "three public expenditures" policy.

Typically, environmental stimuli significantly trigger impulse purchases among consumers by influencing their perceived value and purchase intention, as shown by the stimulus-organism-response (SOR) model (Molinillo et al., 2021). Meanwhile, consumer ethnocentrism, along with perceived behavioral control (PBC) and the subjective norm of the Theory of Planned Behavior (TPB) model, are key factors influencing consumer purchase intentions and behavior (Maksan et al., 2019).

The urgent issues facing Wuliangye can be categorized into two main aspects. Firstly, since Wuliangye's primary customers are dispersed across various regions in China, it's crucial to delve into inherent factors within Chinese consumers themselves. This includes investigating consumer ethnocentrism, subjective norms, perceived value of Wuliangye products, attitudes towards the Wuliangye brand, and purchase intentions. Secondly, given Wuliangye's sales through various channels such as online-to-offline (O2O) models, it's imperative to examine factors related to the sales platforms. This involves studying how environmental stimuli in platform design influence customers' psychological and behavioral purchase decisions, as well as the convenience and security of platform shopping. These factors are integral parts of perceived behavioral control, which in turn impacts customers' psychological and behavioral purchase decisions.

This paper investigates the factors impacting consumer behavior to overcome obstacles in Wuliangye's sales. Through a questionnaire survey in Sichuan Province, it offers strategic recommendations to enhance sales of the Wuliangye liquor series in the Chinese market. To achieve our research objectives, we pose the following research questions (Q1-3):

Q1: How do the urgent issues currently troubling Wuliangye affect the purchasing behavior of Chinese consumers towards the Wuliangye series liquors?

Q2:What is the consumer behavior model of Chinese consumers buying Wuliangye liquor series?

Q3:How can Wuliangye's sales management be more effective in boosting Wuliangye's sales in the Chinese market?

## 2. Research framework and hypothesis development

2.1 Hybrid model stems from ETPB, SOR theory and perceived value theory

The Theory of Planned Behavior (TPB), developed by Ajzen in 1985, has been expanded into the Extended Theory of Planned Behavior (ETPB) by incorporating additional variables such as consumer ethnocentrism (Maksan et al., 2019). The Stimulus-Organism-Response (SOR) model, a framework in psychology and consumer behavior, describes environmental stimuli (S) including information quality, service quality, rewards and recognition, and customization. It integrates perceived value as the organism (O) component, aligning with theories of perceived value (Molinillo et al., 2021). The SOR model, when combined with ETPB, links to purchase behavior or purchase intentions as the response (R) (Maksan et al., 2019).

This article specifically merges ETPB and SOR theory, highlighting perceived value within the SOR framework and proposes a new research model focusing on Sichuan consumer behavior towards Wuliangye, as illustrated in Fig. 1.

2.2. Hypothesis development

Environmental stimuli like information quality, service quality, rewards, recognition, and customization significantly influence customer experiences on social commerce platforms. According to Han et al. (2018), customers particularly appreciate systems that provide rewards and recognition, as these elements motivate engagement and enhance perceived value. The SOR framework highlights the importance of environmental stimuli in shaping consumer behavior, influencing perceptions of value and attitudes towards products, which in turn affect consumer responses (Sultan et al., 2021). Furthermore, Li & Zhang (2022) pinpointed several environmental stimuli that drive online impulsive buying. Based on these insights, this study presents the following hypotheses:

**H1a:** Environmental stimuli positively affect perceived value.

**H1b:** Environmental stimuli positively influence attitudes.

**H1c:** Environmental stimuli positively influence purchase behavior.

Ma et al. (2020) note that consumers who place low importance on achievement often perceive significant economic risks in purchasing foreign products, such as potential job loss and economic restructuring, primarily due to high levels of consumer ethnocentrism that favor local products. Hong et al. (2023) found that consumer ethnocentrism positively influences attitudes by fostering customer engagement. Additionally, Maksan et al. (2019) identified a robust positive correlation between consumer ethnocentrism and the attitudes or purchasing behaviors towards domestic wines in Croatia.

**H2a:** Consumer ethnocentrism positively influences perceived value.

**H2b:** Consumer ethnocentrism positively influences attitudes.

**H2c:** Consumer ethnocentrism positively influences purchase behavior.

Chen (2016) explored how subjective norms and perceived value affect green loyalty in public bike-sharing systems, finding a positive correlation with users' green loyalty. The Theory of Reasoned Action (TRA), introduced by Ajzen & Fishbein (1980), proposes that an individual’s behavior is driven by their behavioral intentions, which are shaped by their attitudes and subjective norms. Furthermore, the TPB, also from Ajzen & Fishbein (1980), suggests that consumers' purchase intentions and perceived behavioral control influence their actual purchasing behaviors. Based on these theories, this paper proposes a series of hypotheses.

**H3a:** Subjective norm positively influences perceived value.

**H3b:** Subjective norm positively influences purchase intention.

**H4a:** Perceived behavioral control positively influences purchase intention.

**H4b:** Perceived behavioral control positively affects purchase behavior.

**H5:** Perceived value positively impacts purchase intention.

**H6:** Attitude positively influences purchase intention.

**H7:** Purchase intention positively impacts purchase behavior.

Based on the aforementioned hypotheses, the following research model (Fig.1) is proposed.

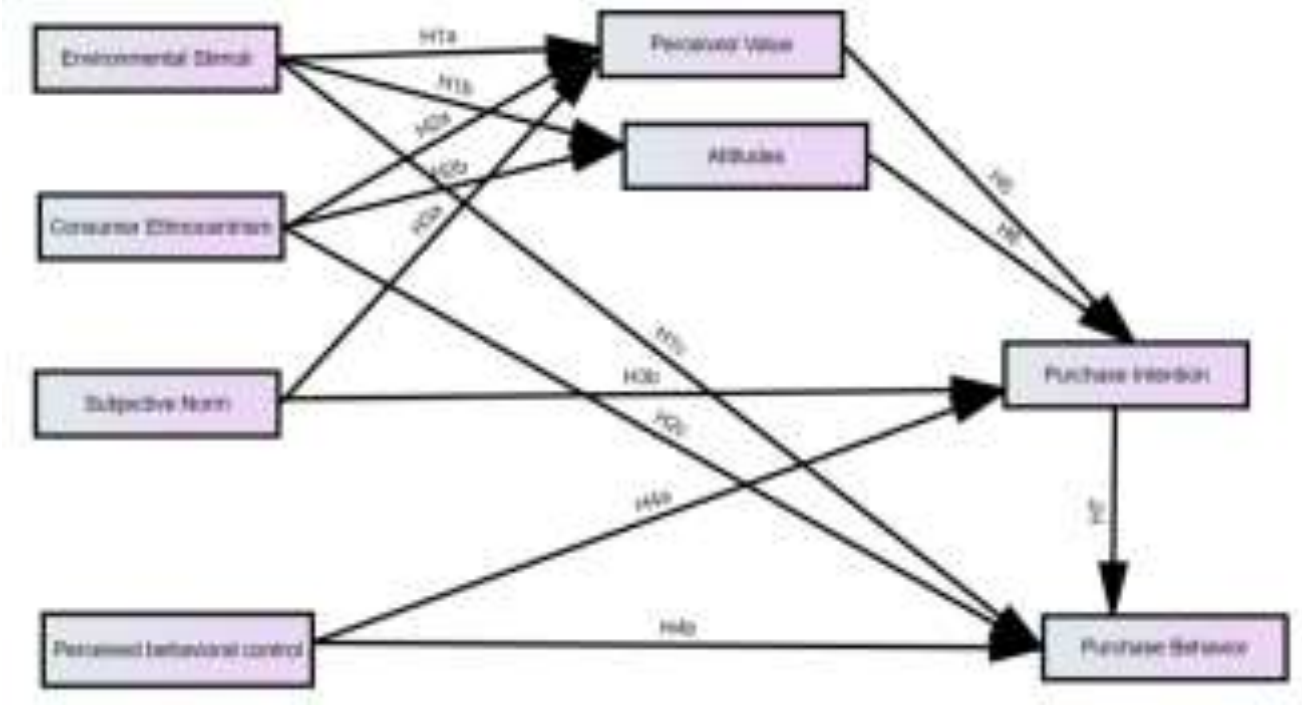

Fig. 1. Research model

Building upon the research model (Fig.1), the paper posits five mediated effect hypotheses as follows.

**H8:** Purchase intention has a mediating effect between perceived behavioral control and purchase behavior (PBC→PurIn→PB)

**H9:** Perceived value and purchase intention have a chain-mediated effect between environmental stimuli and purchase behavior, and it is a partial mediating effect (EnvSt→PerVa→PurIn→PB)

**H10:** Attitudes and purchase intention have a chain-mediated effect between environmental stimuli and purchase behavior, and it is a partial mediating effect (EnvSt→ATT→PurIn→PB)

**H11:** Perceived value and purchase intention have a chain-mediated effect of consumer ethnocentrism on purchase behavior, and it is a partial mediating effect (ConsEth→PerVa→PurIn→PB)

**H12:** Attitudes and purchase intention have a chain-mediated effect of consumer ethnocentrism on purchase behavior, and it is a partial mediating effect (ConsEth→ATT→PurIn→PB).

Note that Hypotheses 9-10 and Hypotheses 11-12 each involve two parallel chain-mediated effects.

## 3. Methodology and results

### 3.1. Data collection and sample

In February 2024, a survey was conducted online, targeting Wuliangye liquor consumers aged 18 and above who had been in the province for at least three years for work or study. Out of 454 responses, 453 were valid. Most respondents were male (71.74%) and young (under 35 years old, 66.45%). The majority had an education level below a master's degree (65.56%) and an above-average income (68.43%). Employees formed the largest group, accounting for 37.31% of the respondents (see Table 1). In the Wuliangye series of liquors, Wuliangye liquor is considered a premium product, while Wuliangchun, Wuliangol, and Wuliangju are mid to low-tier (see Table 2). The survey also shows that Wuliangchun is the most purchased, followed by Wuliangye liquor, and online, Wuliangju leads in sales, again followed by Wuliangye. This indicates the importance of developing mid and low-tier liquors within the Wuliangye series, especially considering that Generation Z, who primarily shop online, prefer these products. Indeed, winning over Gen Z consumers is crucial for future competition in China's liquor industry.

**Table 1 Sample description.**

| n=453 | | N | Percentage |
|---|---|---|---|
| Gender | Male | 325 | 71.74% |
| | Female | 128 | 28.26% |
| Age | 18-35 | 301 | 66.45% |
| | Over 36 | 152 | 33.55% |
| Education | Faculty | 256 | 56.52% |
| | Master/PhD | 156 | 34.44% |
| | Others | 41 | 9.04% |
| Income | Low | 143 | 31.57% |
| | Medium | 177 | 39.07% |
| | High | 133 | 29.36% |
| Work role | Employed | 169 | 37.31% |
| | Unemployed | 155 | 34.21% |
| | Others | 129 | 28.48% |

**Table 2 A survey of consumers' purchase preferences for Wuliangye series liquor**

| | Online or offline purchase | | Only online purchase | |
|---|---|---|---|---|
| n=453 | N | Percentage | N | Percentage |
| Wuliangye liquor | 299 | 66% | 282 | 62.25% |
| Wuliangchun | 303 | 66.89% | 273 | 60.26% |
| Wuliangol | 290 | 64.02% | 266 | 58.72% |
| Wuliangju | 298 | 65.78% | 283 | 62.47% |

3.2. Measures

In this study, consumer purchase behavior is specifically analyzed in terms of frequent purchases of Wuliangye series liquor, encompassing both regular and recent purchase activities as well as a broad range of experiences in buying these products, in line with findings by Maksan et al. (2019). Purchase behavior is distinct from purchase intention as it is influenced by various factors, with perceived behavioral control (PBC) and purchase intention being key predictors of actual purchasing behavior. The measurement of consumer ethnocentrism employs an 8-item version of the CETSCALE, modified from the original 10 items based on input from five management experts and adapted for consumers in Sichuan Province, China, following the approach of Maksan et al. (2019). Environmental stimuli items were also reduced from 10 to 8. A pre-test survey among consumers of Wuliangye liquor in Sichuan yielded 63 valid responses, aimed at testing the clarity and appropriateness of survey questions. Constructs were assessed using a 10-point Likert scale.

### 3.3. Reliability and exploratory factor analysis (EFA)

The results from the SPSS software reveal that Cronbach's α coefficients for all variables were above 0.7, satisfying the established criteria (Cudeck, 2000). The KMO test indicated that all prerequisites were met for structural validity (KMO > 0.7, Bartlett Sig. < 0.01), making the data suitable for exploratory factor analysis (EFA) as referenced in prior studies (Cudeck, 2000). The EFA results show that the 39 identified variables had an information extraction proportion of over 65%, with most exceeding 75%. Consequently, the eight common factors extracted—EnvSt, PerVa, ConsEth, SubN, ATT, PBC, PurIn, and PB—strongly explained all 39 variables. Given that factor loadings for each measurement were above 0.5 and approached 1, while other factors remained below 0.4, the measures demonstrated good discriminant validity. These divisions accurately reflect the factors influencing consumer behavior when purchasing Wuliangye in Sichuan Province.

### 3.4 Confirmatory factor analysis

Firstly, it is evident from Fig. 2 and Table 4 that the fit indices fully meet the requirements. Secondly, the Composite reliability (CR) of each construct in the model exceeds 0.7, aligning with the minimum CR recommended by Malhotra & Dash (2011). Following Fornell & Larcker's (1981) suggestion, internal consistency is also deemed acceptable, as evidenced by all items' outer loadings exceeding 0.7. Lastly, the Average Variance Extracted (AVE) values are not lower than the recommended threshold of 0.5 (Fornell & Larcker, 1981), with consumer ethnocentrism and perceived value exhibiting the highest AVE values (Table 5). Moreover, as indicated in Table 6, the model is validated for discriminant validity, meaning the square root of the AVE of each construct exceeds its other correlation coefficients (Fornell & Larcker, 1981).

### 3.5 SEM and hypotheses testing (H1-H7)

Fig. 3 and Table 7 demonstrate that the structural equation model fits well, with all fit indices meeting the standard values. In fact, the standard values encompass three aspects: S.E. should be close to 0; C.R. should be greater than 1.96 to achieve statistical significance; P-values should generally be less than 0.05, indicating that the regression coefficients are significant at a 95% confidence level, where S.E. stands for the standard error of the regression weights, indicating the uncertainty in the estimation of regression coefficients. C.R. denotes the critical ratio of the regression weights, used to test the significance of regression coefficients. P-values indicate the significance level of the regression coefficients, typically used to test whether there is a significant relationship between the predictor and outcome variables (Malhotra &

Dash, 2011; Fornell & Larcker, 1981). Therefore, we confirm from Table 8 that all the hypotheses (H1-H7) are supported.

3.6 Mediation effects versus chain-mediated effects hypothesis testing (H8-H12)
After analyzing the AMOS structural equation model (refer to Fig. 3), the standardized estimates for mediation effects were presented in Table 9. These results indicate support for all hypotheses concerning mediating effects, especially H8, as evidenced by the significance of total, direct, and indirect effects (Qiu et al., 2023).

To examine chain-mediated effects, we defined M1 = H1a * H5 * H7, M2 = H1b * H6 * H7, M3 = H2a * H5 * H7, and M4 = H2b * H6 * H7. Table 10 shows that all four chain-mediated effects are partial mediations. In summary, all mediating hypotheses (H8-H12) are upheld (Qiu et al., 2023).

Additionally, Table 10 reveals that the comparison of parameter magnitudes signifies the precedence of the path "ConsEth→PerVa→PurIn→PB" over "ConsEth→ATT→PurIn→PB," highlighting the perceived significance of product value over brand attitudes.

## 4. Discussion

The study utilized an online survey of consumers from Sichuan Province, China. Findings indicate that the ETPB model within the SOR framework effectively predicts the intentions and behaviors of Chinese consumers regularly purchasing Wuliangye products. Key contributions of this study include insights that can help formulate marketing strategies to boost domestic sales of Wuliangye in China, providing rationalized recommendations for Wuliangye's sales management department. The research confirmed that intention does not always translate directly into behavior, with Perceived Behavioral Control and purchase intention both playing a significant role in consumer purchasing actions, echoing findings by Karki & Hubacek (2015), and Maksan et al. (2019). Environmental stimuli emerged as a strong influencing factor of consumer behavior, consistent with findings by Molinillo et al. (2021). The findings emphasize the importance of developing the convenience of online sales platforms and establishing membership programs for consumers. Besides, consumer ethnocentrism, added as an additional construct, proved to be a robust predictor of attitudes towards and behaviors regarding Wuliangye purchases, aligning with findings by Maksan et al. (2019). Moreover, the parallel chain-mediated effect analysis highlighted that perceived product value slightly surpasses brand attitudes, underscoring the importance of the presenter's product expertise in marketing campaigns. Finally, statistical analysis indicated that online sales dominate Wuliangye's sales channels, particularly among Generation Z, with Wuliangju, a mid-to-low brand, being the top seller online, which underscores the importance of developing mid to mid-to-low product lines to attract Gen Z consumers.

## 6. Conclusion and limitations

This study shows that a combined model of ETPB and SOR theory is effective and applicable in predicting regular purchases of Wuliangye liquor by consumers. Environmental stimuli and consumer ethnocentrism strongly influence attitudes and behaviors towards regular purchases of Wuliangye, and they partially affect consumer behavior through chain-mediated effects involving perceived value and purchase intention, or attitude and purchase intention. Additionally, Perceived Behavioral Control and purchase intention both are significant predictors of buying behavior.

This article presents preliminary findings on the consumer behavior towards Wuliangye, a well-known liquor from Sichuan Province, China. It outlines several limitations due to time, cost, and researcher capabilities. The study acknowledges incomplete exploration of all factors affecting consumer preferences for Wuliangye and suggests that future research should widen this scope. Additionally, there were issues with questionnaire design, especially terms related to consumer ethnocentrism that may not resonate with Chinese linguistic norms, risking measurement inaccuracies. The research was also limited to consumers from Sichuan, recommending future studies to include a broader geographical sample to better represent and validate the findings.

**Table 3 Reliability.**

| Variables | No. | Measurement Items | α |
|---|---|---|---|
| **ConsEth** | CE1 | Only those products that are unavailable in the China should be imported. | **0.94** |
| | CE2 | Chinese products, first, last and foremost. | |
| | CE3 | It is not right to purchase foreign products, because it puts Chinese people out of jobs. | |
| | CE4 | A real Chinese person should always buy the products made in China | |
| | CE5 | We should purchase products manufactured in China instead of letting other countries get rich off us. | |
| | CE6 | Chinese people should not buy foreign products, because this hurts Chinese business and causes unemployment. | |
| | CE7 | It may cost me in the long run but I prefer to support Chinese products. | |
| | CE8 | We should buy from foreign countries only those products that we cannot obtain within our own country. | |
| **EnvSt** | ES1 | This social commerce site of Wuliangye provides me with the precise information I need | **0.92** |
| | ES2 | The information content provided by the Wuliangye social commerce site meets my needs | |
| | ES3 | I think the information content provided by the Wuliangye social commerce site is reliable | |
| | ES4 | The Wuliangye social commerce site provides me with up-to-date information | |
| | ES5 | When I have a problem, the Wuliangye social commerce site service shows a sincere interest in solving it | |
| | ES6 | The Wuliangye social commerce site service is always willing to help me | |
| | ES7 | I feel safe in my transactions with the Wuliangye social commerce site service in terms of security and privacy protection | |
| | ES8 | The Wuliangye social commerce site service has the knowledge to answer my questions | |
| **SubN** | SN1 | My family thinks I should buy Wuliangye-series liquor. | **0.78** |
| | SN2 | My friends think I should buy Wuliangye-series liquor. | |
| | SN3 | News and magazines affect my purchase decisions of Wuliangye-series liquor. | |
| **PBC** | PBC1 | I have enough opportunity in making purchase decision | **0.78** |
| | PBC2 | I have the capacity to make a purchase decision | |
| | PBC3 | I have enough control in my purchase decision | |
| | PBC4 | I have enough knowledge to make purchase decision | |
| **PerVa** | PV1 | Considering the money I pay for buying products on this social commerce site or physical store, internet shopping or offline purchasing here represents a good deal | **0.86** |
| | PV2 | Considering the effort I make in shopping at this social commerce site or physical store, internet shopping or offline purchasing here is worthwhile | |
| | PV3 | Considering the risk involved in shopping at this social commerce site or physical store, internet shopping or offline | |
| | | purchasing here is of value | |
| **ATT** | ATT1 | Regular buying of Wuliangye-series liquor is pleasure for me. | **0.78** |
| | ATT2 | Regular buying of Wuliangye-series liquor evokes positive emotions in me. | |
| | ATT3 | Buying a Wuliangye-series liquor is complete ritual for me. | |
| | ATT4 | Buying a Wuliangye-series liquor is funny. | |
| **PurIn** | PI1 | I intend to buy Wuliangye-series liquor regularly. | **0.79** |
| | PI2 | I have a plan of buying a Wuliangye-series liquor within 1 year | |
| | PI3 | Next time I browse liquor website or visit a physical store to buy liquor, I am more inclined to purchase a Wuliangye-series liquor. | |
| | PI4 | I have commitment to buy a Wuliangye-series liquor | |
| **PB** | PB1 | I am buying Wuliangye-series liquor regularly. | **0.81** |
| | PB2 | In my shopping basket is regularly Wuliangye-series liquor. | |
| | PB3 | When I am buying liquor, I regularly choose Wuliangye-series liquor. | |
| | PB4 | In the past ten months, I have bought Wuliangye-series liquor. | |
| | PB5 | I have had many experiences of buying Wuliangye-series liquor. | |

Note. Adapted from SPSS Software Result. EnvSt=Environmental Stimuli, PerVa=Perceived Value, ConsEth=Consumer Ethnocentrism, SubN=Subjective Norm, ATT=Attitudes,PBC=Perceived behavioral control, PurIn=Purchase Intention, PB= Purchase behavior. **α** represents Cronbach's **α**.

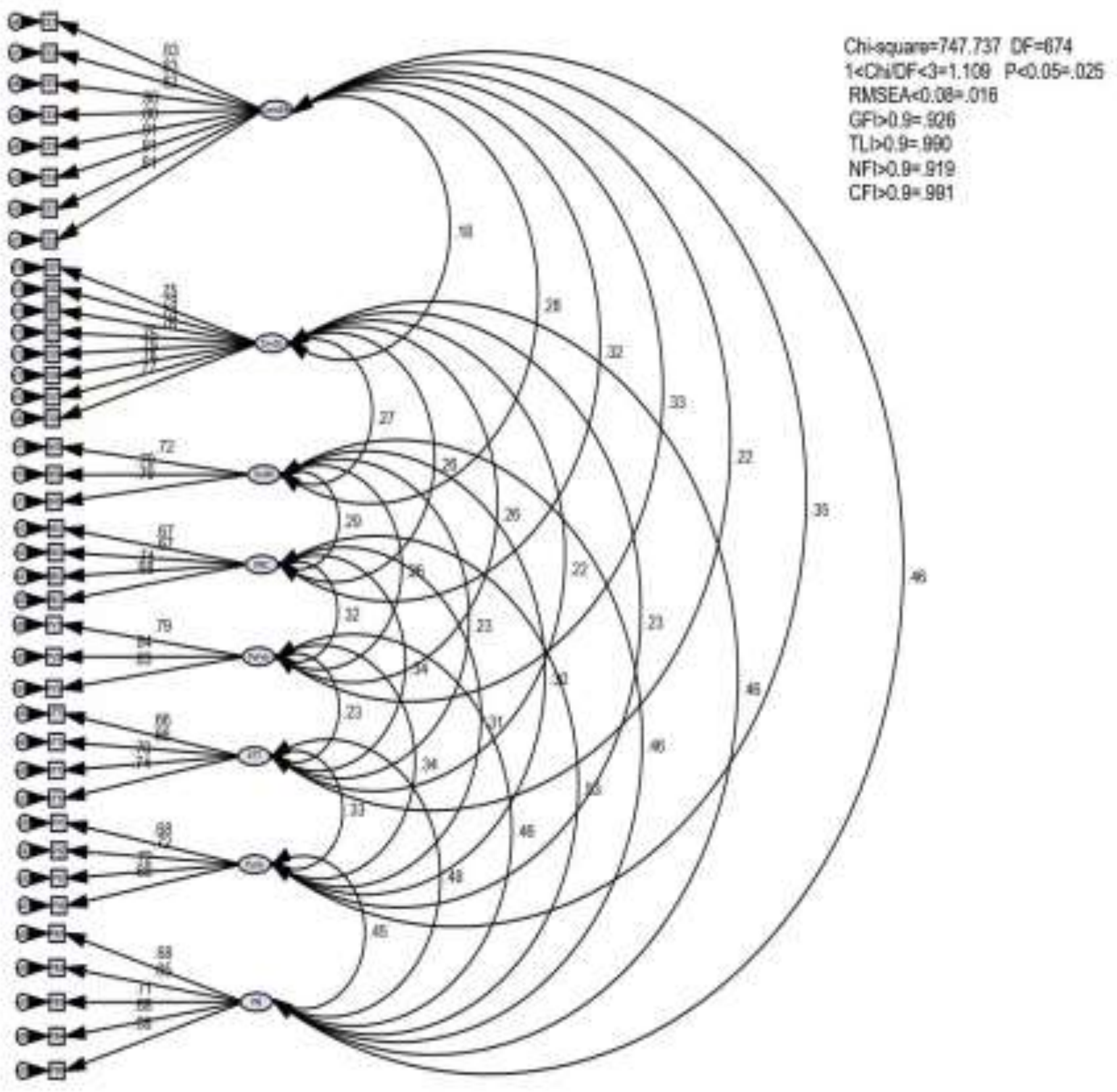

**Fig. 2 CFA with a title**

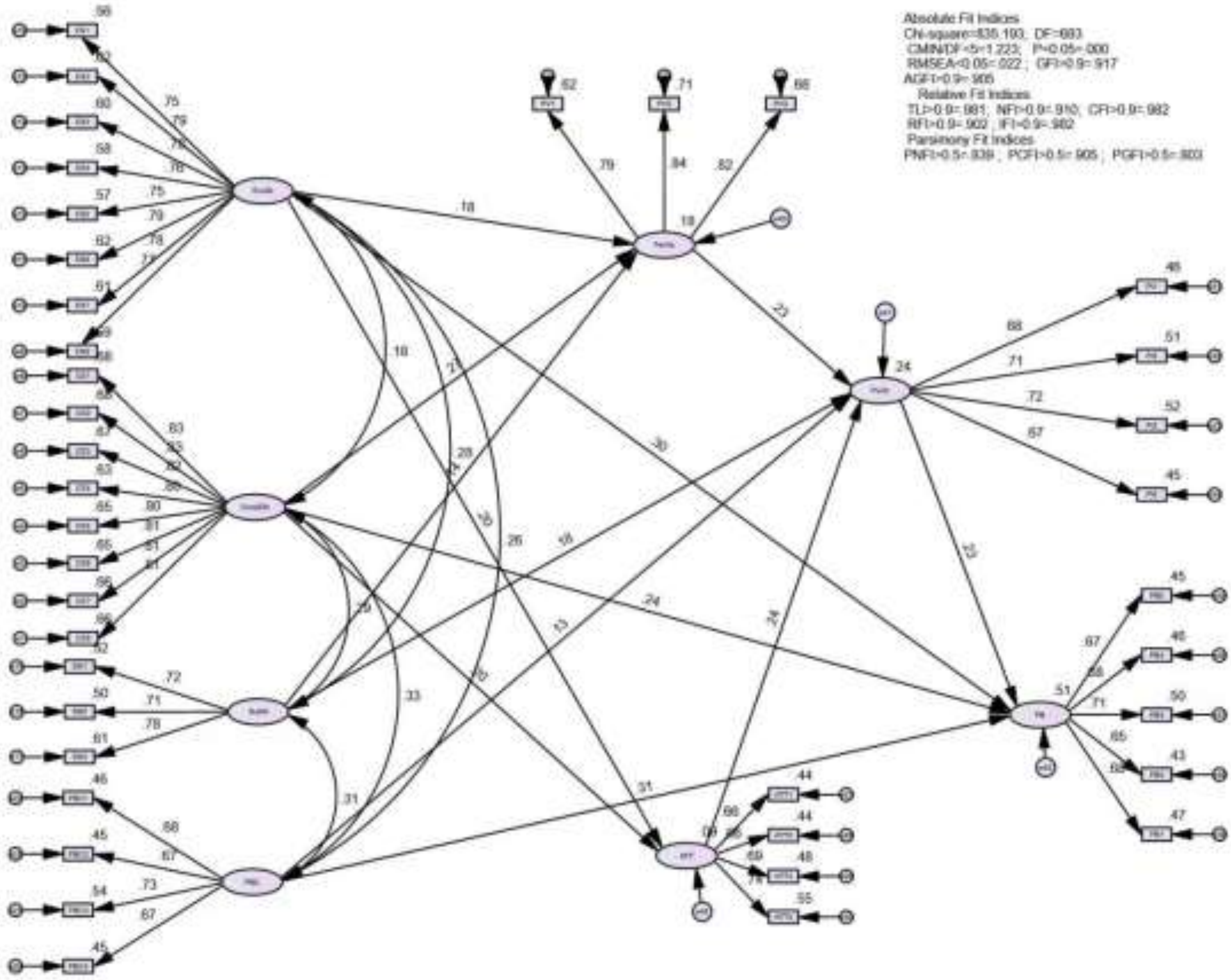

**Fig. 3 SEM with a title**

**Table 4 Model fit coefficients from CFA**

| Item | Chi/DF | P | RMSEA | GFI | AGFI | TLI | NFI | CFI | PNFI | PGFI |
|---|---|---|---|---|---|---|---|---|---|---|
| Value | 1.109 | 0.025 | 0.016 | 0.926 | 0.914 | 0.990 | 0.919 | 0.991 | 0.836 | 0.800 |
| Criteria | < 5 | < 0.05 | < 0.05 | > 0.9 | > 0.9 | > 0.9 | > 0.9 | > 0.9 | > 0.5 | > 0.5 |

**Table 5 Convergent validity.**

| **Path** | CE(i)←ConsEth,i=1,2,...,8 | ES(i)←EnvSt,i=1,...,8 | SN(i)←SubN, i=1,2,3 | PBC(i)←PBC, i=1,2,3,4 | PV(i)←PerVa, i=1,2,3 | ATT(i)←ATT, i=1,2,3,4 | PI(i)←PurIn, i=1,2,3,4 | PB(i)←PB, i=1,...,5 |
|---|---|---|---|---|---|---|---|---|
| **CR** | 0.9401 | 0.9212 | 0.7835 | 0.7838 | 0.8586 | 0.7848 | 0.7953 | 0.813 |
| **AVE** | 0.7 | 0.6 | 0.5 | 0.5 | 0.7 | 0.5 | 0.5 | 0.5 |

**Table 6 Discriminant validity**

| **Variable** | **PB** | **PurIn** | **ATT** | **PerVa** | **PBC** | **SubN** | **EnvSt** | **ConsEth** |
|---|---|---|---|---|---|---|---|---|
| PB | **0.7071** | | | | | | | |
| PurIn | 0.447** | **0.7071** | | | | | | |
| ATT | 0.479** | 0.334** | **0.7071** | | | | | |
| PerVa | 0.457** | 0.344** | 0.233** | **0.8367** | | | | |
| PBC | 0.532** | 0.313** | 0.336** | 0.324** | **0.7071** | | | |
| SubN | 0.462** | 0.302** | 0.233** | 0.262** | 0.291** | **0.7071** | | |
| EnvSt | 0.463** | 0.23** | 0.216** | 0.258** | 0.256** | 0.275** | **0.7746** | |
| ConsEth | 0.459** | 0.346** | 0.22** | 0.33** | 0.324** | 0.277** | 0.18** | **0.8367** |
| **AVE** | 0.5 | 0.5 | 0.5 | 0.7 | 0.5 | 0.5 | 0.6 | 0.7 |

Note. ** was significantly correlated at 0.01 level. The value in bold on the upper right corner is the square root of AVE. Here $0.7071^2=0.5$, $0.7746^2=0.6$, $0.8367^2=0.7$.

**Table 7 Model Fit Summary of the SEM**

| Ite | CMI | P | GF | A | | TL | NF | RF | IFI | TL | CF | PN | PC | RM |
|---|---|---|---|---|---|---|---|---|---|---|---|---|---|---|

| m | N/DF | | I | GFI | PGFI | I | I | I | | I | I | FI | FI | SEA |
|---|---|---|---|---|---|---|---|---|---|---|---|---|---|---|
| Value | 1.223 | 0.000 | 0.917 | 0.905 | 0.803 | 0.981 | 0.910 | 0.902 | 0.982 | 0.981 | 0.982 | 0.839 | 0.905 | 0.022 |
| Criteria | < 5 | < 0.05 | > 0.9 | > 0.9 | > 0.5 | > 0.9 | > 0.9 | > 0.9 | > 0.9 | > 0.9 | > 0.9 | > 0.5 | > 0.5 | < 0.05 |

**Table 8 Regression weights**

| Path | PerVa←ConsEth | ATT←ConsEth | PerVa←EnvSt | ATT←EnvSt | PerVa←SubN | PurIn←PerVa | PurIn←SubN | PurIn←PBC | PurIn←ATT | PB←PBC | PB←PurIn | PB←EnvSt | PB←ConsEth |
|---|---|---|---|---|---|---|---|---|---|---|---|---|---|
| Estimate | 0.267 | 0.203 | 0.179 | 0.196 | 0.144 | 0.234 | 0.178 | 0.126 | 0.238 | 0.311 | 0.231 | 0.299 | 0.24 |
| S.E. | 0.048 | 0.039 | 0.053 | 0.045 | 0.069 | 0.045 | 0.057 | 0.064 | 0.058 | 0.055 | 0.047 | 0.034 | 0.03 |
| C.R. | 5.023 | 3.674 | 3.395 | 3.508 | 2.442 | 4.051 | 2.851 | 2.09 | 4.049 | 5.356 | 4.382 | 6.001 | 4.878 |
| P | *** | *** | *** | *** | 0.015 | *** | 0.004 | 0.037 | *** | *** | *** | *** | *** |
| Label | H2a | H2b | H1a | H1b | H3a | H5 | H3b | H4a | H6 | H4b | H7 | H1c | H2c |

**Table 9 Total effect, Direct effect and Indirect effect**

| Standardized estimate of H8 : PBC→PurIn→PB | | | | | | | | |
|---|---|---|---|---|---|---|---|---|
| Total effect | Lower bound | Upper bound | Direct effect | Lower bound | Upper bound | Indirect effect | Lower bound | Upper bound |
| 0.340 | 0.239 | 0.447 | 0.311 | 0.204 | 0.416 | 0.029 | 0.003 | 0.069 |
| Standardized estimate of the mediation effect of EnvSt on PB | | | | | | | | |
| Total effect | Lower bound | Upper bound | Direct effect | Lower bound | Upper bound | Indirect effect | Lower bound | Upper bound |
| 0.319 | 0.146 | 0.300 | 0.299 | 0.198 | 0.390 | 0.020 | 0.007 | 0.044 |
| Standardized estimate of the mediation effect of ConsEth on PB | | | | | | | | |
| Total effect | Lower bound | Upper bound | Direct effect | Lower bound | Upper bound | Indirect effect | Lower bound | Upper bound |
| 0.265 | 0.164 | 0.363 | 0.240 | 0.142 | 0.343 | 0.026 | 0.008 | 0.052 |

**Table 10. User-defined estimands**

| Parameter | Estimate | Lower | Upper | P |
|---|---|---|---|---|
| M1 | 0.007 | 0.002 | 0.017 | 0.000 |
| M2 | 0.007 | 0.002 | 0.019 | 0.002 |
| M3 | 0.009 | 0.003 | 0.02 | 0.001 |
| M4 | 0.007 | 0.002 | 0.019 | 0.001 |